\documentclass[letterpaper,twocolumn,fleqn]{article} 

\usepackage{ist}
\usepackage{amsmath,graphicx}
\usepackage{epstopdf} 
\usepackage{amsmath}
\usepackage[flushleft]{threeparttable}

\def\etal{{\textit{et al}. }}
\newcommand{\imheight}{.85in}
\newcommand{\imwidth}{.98in}

\graphicspath{
{./figures/}
}

\title{A Coarse-To-Fine Framework For Video Object Segmentation}

\author{ Chi Zhang\textsuperscript{1}, Rochester Institute of Technology, Rochester, NY 14623, USA \\
Alexander Loui, Kodak Alaris Imaging R\&D, Rochester, NY 14615, USA}

\date{} 

\begin{document} 

\maketitle 

\thispagestyle{empty} 


\begin{abstract}
In this study, we develop an unsupervised coarse-to-fine video analysis framework and prototype system to extract a salient object in a video sequence. This framework starts from tracking grid-sampled points along temporal frames, typically using KLT tracking method. The tracking points could be divided into several groups due to their inconsistent movements. At the same time, the SLIC algorithm is extended into 3D space to generate supervoxels. Coarse segmentation is achieved by combining the categorized tracking points and supervoxels of the corresponding frame in the video sequence. Finally, a graph-based fine segmentation algorithm is used to extract the moving object in the scene. Experimental results reveal that this method outperforms the previous approaches in terms of accuracy and robustness. 
\end{abstract}

\footnotetext[1]{Work performed at Kodak Alaris during an internship from Rochester Institute of Technology.}


\section{Introduction}
\label{sec:introduction}

Object level video segments are semantically meaningful spatiotemporal units such as moving persons, moving vehicles, flowing river, etc. Segmentation of video sequence into a number of component regions would benefit many higher level vision based applications such as scene analysis, object localization and content understanding. However, single target object extraction would be a more demanding task considering consumer’s needs. In many cases, a consumer video sequence simply targets at capturing a single object's movement in a specific environment such as dancing, skiing, running, etc. In general, motion object detection and extraction for a static video camera is relatively straightforward since the background barely changes and a simple frame differencing would be able to extract a moving foreground object. However, it is still challenging for the object moving on a cluttered and/or dynamic background.

The goal of background modeling and foreground object extraction is to build a model of the background/foreground in an offline manner and extract the object of interest by comparing the estimated model with the frames. The model must be robust enough to cope with background changes in different ways. In recent years, a trend towards modeling spatio-temporal uniform (in terms of either appearance or motion) regions instead of single pixels has been observed \cite{Lim_14}. These works rely on superpixels/supervoxels for object segmentation in videos. However, these methods is computationally expensive and group superpixels together according to pure spatio-temporal similarity without exploiting real-world object features. As an improvement, Giordano \etal \cite{Giordano_15} proposed an approach without making any specific assumptions about the videos and it relies on how objects are perceived by humans according to Gestalt laws. Khoreva \etal \cite{Khoreva_15} proposed an empirical approach to learn both the edge topology and weights of the graph. The most confident edges are selected by the graph structure while the classifiers are learned to combine features and seamlessly integrated by its accuracy. In \cite{Papon_13} and \cite{Trulls_14}, FPFH and HoG have been used as features to represent superpixels. The high dimension feature space slows down the computation, although some improvements (e.g., \cite{Neubert_14}) were proposed to provide a better balance of trade off between segmentation quality and runtime.

Moreover, much research has been devoted to graph models for segmentation, such as \cite{Khoreva_15} and \cite{Galasso_14}. Fan and Loui \cite{Fan_15} proposed a graph-based approach that models the data in a feature space, which emphasizes the correlation between similar pixels while reducing the inter-class connectivity between different objects. In \cite{Li_15}, a reduced superpixel graph was reweighted such that the resulting segmentation was equivalent to the full graph under certain assumptions.
 
In this work, we develop a novel coarse-to-fine framework and prototype system for automatically segmenting a video sequence and extracting a salient moving object from it. The proposed framework comprises of point tracking and motion clustering of pixels into groups. In parallel, a pixel grouping method is used to generate supervoxels for the corresponding frames of the video sequence. Coarse segmentation is achieved by combining the results of previous steps. Subsequently, a graph-based technique is used to perform fine segmentation and extraction of the salient object. The following section presents the proposed coarse-to-fine video segmentation framework and the details of the key component algorithms. Then the performance evaluations and experimental results will be discussed in an individual section. Finally, some concluding remarks are presented in the last Section.

\section{System Framework and Algorithms}
\label{sec:algorithm}

\begin{figure}[htb]
\begin{minipage}[b]{1.0\linewidth}
  \centering
  \centerline{\includegraphics[width=8.5cm]{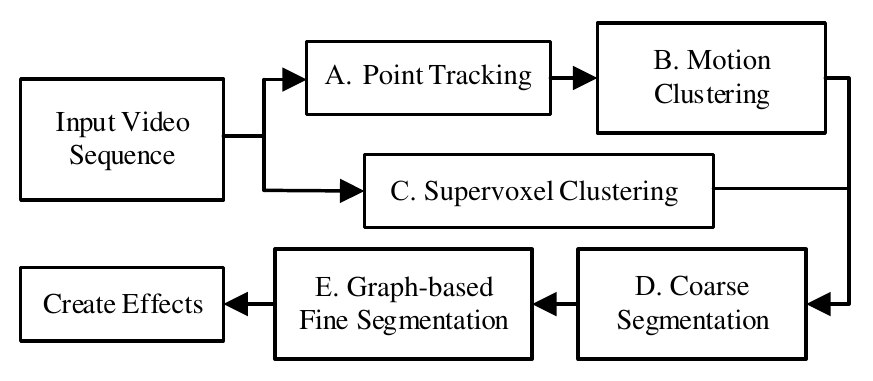}}
\end{minipage}
\caption{The overall framework of the proposed algorithm.}
\label{fig:flowchart}
\vspace{-.1in}
\end{figure}

The proposed framework is shown in Fig. \ref{fig:flowchart}, and consists of several stages: 1) the point tracking algorithm is applied to the consecutive frames of the input video, and then 2) these tracking points are clustered into groups; in parallel, 3) a pixel grouping method is used to generate supervoxels for the corresponding frame of the video sequence; 4) coarse segmentation is achieved by combining the results of previous steps; finally, 5) a graph-based segmentation technique is used to perform fine segmentation and generate a mask of most salient object.

This video segmentation scheme exhibits state-of-the-art boundary adherence, improves the performance of segmentation algorithms with reduced memory consumption. This new approach is a major enhancement to the previous graph-based framework \cite{Fan_15}, with the following distinctions and advantages:
\begin{itemize}
  \item We deal with the video sequence with any resolution and any length, i.e., there is no restriction on the size of the video. For a long video sequence, it is divided into small clips that are processed by the system one by one.
  \item The parallel approach combines the spatial and temporal information and takes advantage of both graph-based algorithms and pixel grouping methods. Consequently, it provides a marked improvement on accuracy and speed.
  \item It is an unsupervised scheme, i.e., there is no user interaction required to generate the accurate object mask.
\end{itemize}

\subsection{A. Points Tracking}
\label{ssec:KLT}

\begin{figure}[htb]
\begin{minipage}[b]{.32\linewidth}
  \centering
  \centerline{\includegraphics[width=\imwidth, height=\imheight]{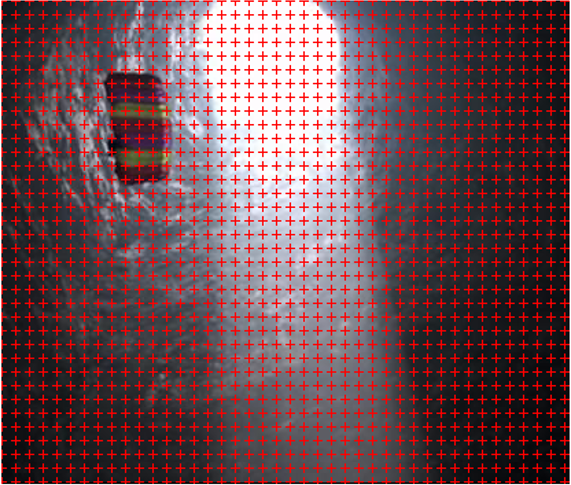}}
  \centerline{(a)}\medskip
\end{minipage}
\begin{minipage}[b]{.32\linewidth}
  \centering
  \centerline{\includegraphics[width=\imwidth, height=\imheight]{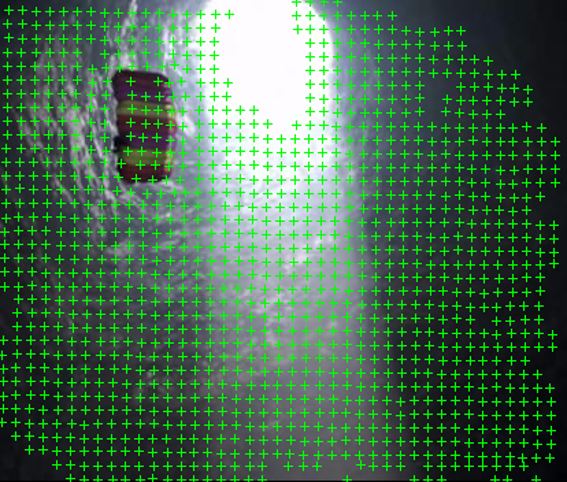}}
  \centerline{(b)}\medskip
\end{minipage}
\begin{minipage}[b]{.32\linewidth}
  \centering
  \centerline{\includegraphics[width=\imwidth, height=\imheight]{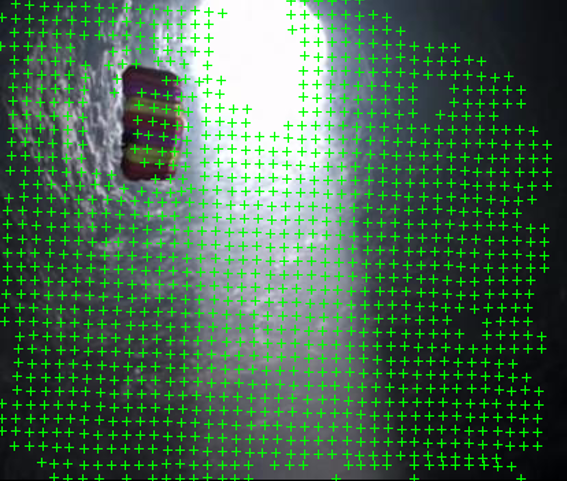}}
  \centerline{(c)}\medskip
\end{minipage}
  \vspace{-.1in}
\caption{KLT point tracking. (a) Selected tracking points in the 1st frame; (b) and (c) Tracking points in the 3rd and 5th frame. (Please print in color.)}
\label{fig:KLT}
\end{figure}

There are a lot of widely-used points tracking algorithms, such as particle filtering \cite{Khoreva_14} and mean shift tracking \cite{Comaniciu_02}, and each of them has its own characteristics. A popular and well-performed video object tracking algorithm is the Kanade-Lucas-Tomasi (KLT) point tracker \cite{Fu_15,Lucas_81}. The algorithm basically provides the trajectories of a bundle of points. In our work, the points to be tracked are selected in a grid-based manner in order to make the initial points distributed uniformly in the entire frame, as shown by the red dots in Fig. \ref{fig:KLT}(a). As the point tracking algorithm progresses over time, points can be lost due to lighting variation, out of plane rotation, or articulated motion as shown in Fig. \ref{fig:KLT}(b) and Fig. \ref{fig:KLT}(c). To track an object over a long period of time, we may need to reacquire points periodically.

There are some algorithms proposed to improve the accuracy of KLT points tracking. One such proposal is the TLD algorithm proposed by Kalal \cite{KALAL_10}.

The KLT points tracker requires some premises: 1) the luminance between two adjacent frames should be constant; 2) the object moves continuously in time domain, otherwise the movement should be ``small'' enough; 3) a point and its neighborhood have similar motion vector, i.e., spatial consistent. Intuitively, if a window $w$ in frame $I$ is the same as that in the adjacent frame $J$, we have $I(x,y,t)=J(x',y',t+\tau)$. The constant-luminance hypothesis holds the equality and gets rid of the effect of luminance changes. The second premise ensures the existence of the tracking points. The points in the same window that have the same offset is guaranteed by the third premise.

\subsection{B. Motion Clustering}
\label{ssec:SSC}

In a video sequence, the collection of points which locates in the high-dimensional space often lie close to low-dimensional structures corresponding to several classes the data belongs to. The Sparse Subspace Clustering (SSC) algorithm proposed by Elhamifar and Vidal \cite{Y_Zhang_15} clusters tracking points that lie in a union of low-dimensional subspaces. The point trajectories acquired by KLT point tracker are grouped into two clusters using SSC algorithm. Among infinitely many possible representations of the data in terms of other points, a sparse representation corresponds to selecting a few points from the same subspace. This motivates solving a sparse optimization problem whose solution is used in a spectral clustering framework to infer the clustering of data into subspaces. 

\begin{figure}[htb]
\begin{minipage}[b]{.5\linewidth}
  \centering
  \centerline{\includegraphics[width=.9\linewidth]{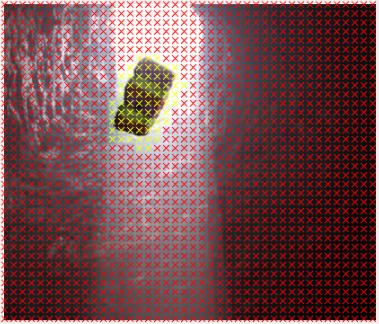}}
\end{minipage}
\begin{minipage}[b]{.5\linewidth}
  \centering
  \centerline{\includegraphics[width=.9\linewidth]{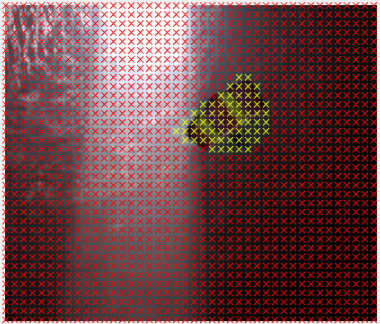}}
\end{minipage}
\caption{Demonstrations of point trajectory clustering using SSC on two frames. The yellow and red markers represent two clusters, foreground and background respectively. (Please print in color.)}
\label{fig:SSC}
\end{figure}

Fig. \ref{fig:SSC} shows the clustering results on two frames. Due to the fact that the object moves in a different way from the background does, the tracking points on the object are separated from the points on the background. Actually, this algorithm can be solved efficiently and can handle data points near the intersections of subspaces. Another key advantage of this algorithm with respect to the state-of-the-art is that it can deal with data nuisances, such as noise, sparse outlying entries, and missing entries, directly by incorporating the model of the data into the sparse optimization program.

\subsection{C. Supervoxel Clustering}
\label{ssec:SLIC}

In our work, the Simple Linear Iterative Clustering (SLIC) \cite{Achanta_12} is extended to 3D space for dealing with 3D data clustering problem.

Considering the aspect of computational efficiency, the entire video sequence is divided into clips and each chip contains a fixed number of frames, which is determined by the computing ability of the processor. Each clip can then be processed individually. The resolution of consumer videos is sometimes comparable or higher than 720p HD videos, which contain too much details in each frame and cause undesired effects and redundant computations on the 3D SLIC performance. Bilateral filtering \cite{He_13} can be used on each frame in order to solve this problem so that the edges around the objects are preserved and the other regions are smoothed. Also, bilateral filtering reduces the noise in each channel.

Suppose that the desired number of supervoxels on each frame is $n$ and the thickness of each supervoxel is $D$ along the temporal axis. Assuming that the supervoxels are initially square in each frame and approximately equal-sized. All cluster centers are initialized by sampling the clip on a regular grid spaced $S$ pixel apart inside each frame and $t$ pixel between frames (along temporal axis). Without considering the accuracy for small color differences, the video sequence is converted into CIELAB space, since the nonlinear relation for $L^*$, $a^*$, and $b^*$ good model to mimic the nonlinear response of the eye. Furthermore, uniform changes of components in the CIELAB color space aim to correspond to uniform changes in perceived color, so the relative perceptual differences between any two colors can be approximated by treating each color as a point in a three-dimensional space and taking the Euclidean distance between them. Also, the motion information can be represented by motion vectors obtained from optical flow. Consequently, each cluster is then represented by the vector
\begin{align}
C=[x\ y\ z\ L^*\  a^*\  b^*\  u\ v]
\end{align}
where $x$ and $y$ represent the spatial location and $z$ carries the temporal information, $L^*$,$a^*$ and $b^*$ represent the spectral information and $u$,$v$ are motion information extracted by optical flow.

\begin{figure}[htb]
  \centering
\begin{minipage}[b]{.85\linewidth}
  \centerline{\includegraphics[width=.9\linewidth]{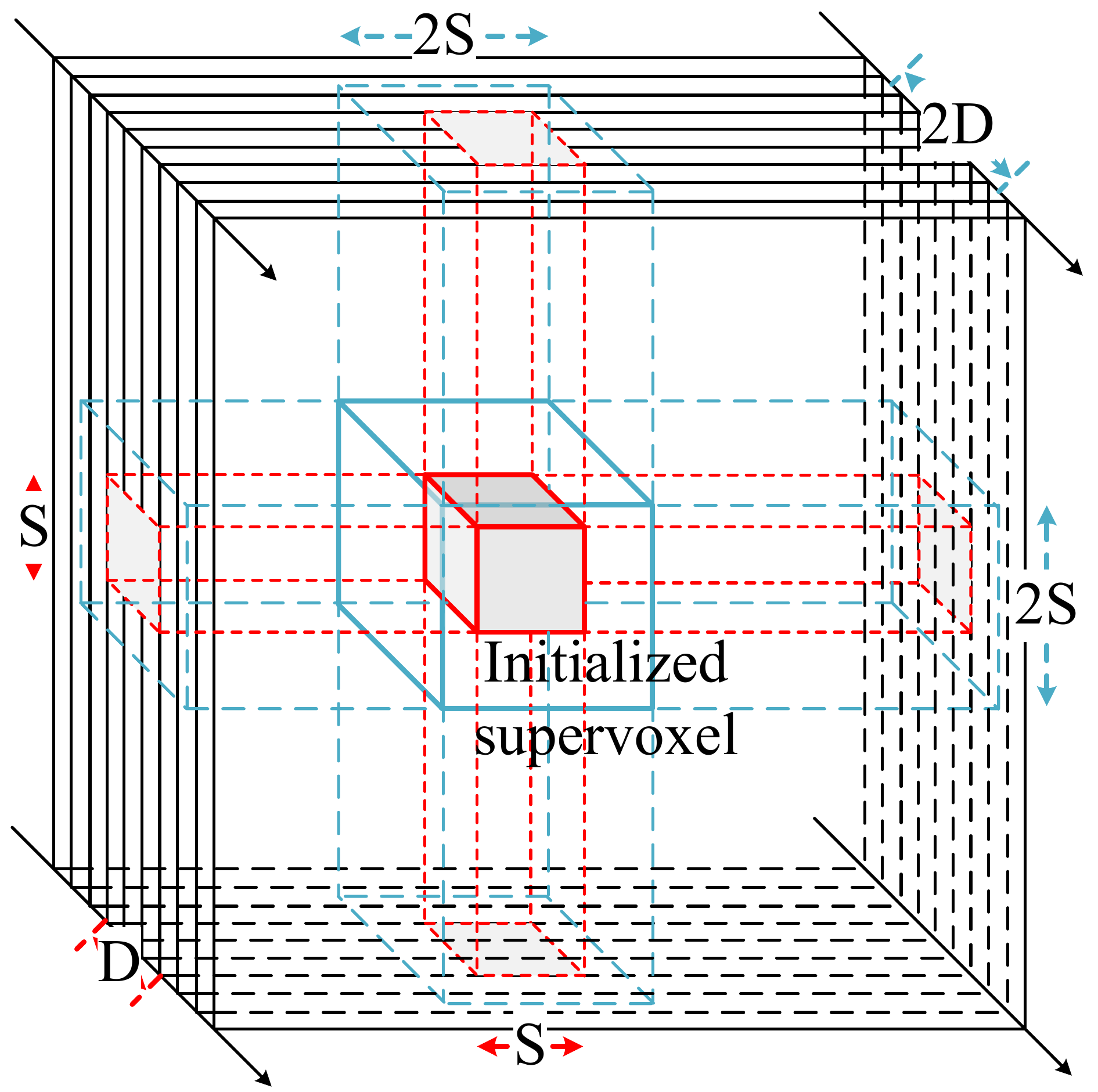}}
\end{minipage}
\caption{Initialization and the search region of supervoxel. Red box shows the initialized supervoxel along $D$ consecutive frames. Blue box is the searching area for this cluster. Each pixel is calculated eight times since it enclosed by eight cluster search region. (Please print in color.)}
\label{fig:supervoxel}
\end{figure}

In the assignment step, the cluster of each pixel is determined by calculating the distance between the pixel itself and the cluster center in the $2S\times 2S\times 2D$ search region, as shown in Fig. \ref{fig:supervoxel}.

The problem arises when the distance is measured. In this case, the distances in each domain are calculated separately and then combined after multiplying the appropriate weights, i.e., the distance $d$ is defined by the pixel location, the CIELAB color space and motion vector in the image is as follows:
\begin{align}
d=\sqrt{\dfrac{d_l^2}{2S^2+D^2}+\dfrac{d_c^2}{m}+\dfrac{w_m\cdot d_m^2}{RS}}
\end{align}
where $m$ is the regularity that controls the compactness of the supervoxel, $w_m$ is a weight on motion information, $R$ is frame rate, and 
\begin{align}
d_l &= \sqrt{\Delta x^2+\Delta y^2+w_z\cdot \Delta z^2}\\
d_c &= \sqrt{w_{L^*}\cdot \Delta L^{*2}+\Delta a^{*2}+\Delta b^{*2}}\\
d_m &= \sqrt{\Delta u^2+\Delta v^2}=\sqrt{\Delta\dot{x}^2+\Delta\dot{y}^2}
\end{align}
where $w_z$ and $w_{L^*}$ are the weights for the temporal distance and $L^*$ channel. In the distance measure, the location is normalized by the maximum distance in the 3D lattice $2S^2+D^2$ according to Fig. \ref{fig:supervoxel}.  The weight for the depth component $w_z$ is introduced since the inter-frame (lateral) position distance should be treated differently as in-frame (transverse) distance. Considering two adjacent supervoxels with depth $D$ in the temporal axis, these two supervoxels would shrink transversely and expand up to 2D in lateral direction during the iterations if the region surrounded is relatively uniform and the weight $w_z$ is small. This causes the increased number of clusters on a single frame, which is unexpected for some applications.

\begin{figure}[htb]
\begin{minipage}[b]{.32\linewidth}
  \centering
  \centerline{\includegraphics[width=\imwidth, height=\imheight]{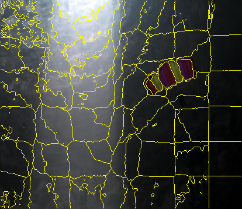}}
\end{minipage}
\begin{minipage}[b]{.32\linewidth}
  \centering
  \centerline{\includegraphics[width=\imwidth, height=\imheight]{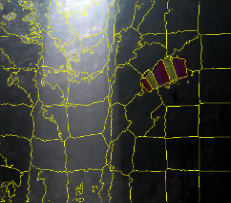}}
\end{minipage}
\begin{minipage}[b]{.32\linewidth}
  \centering
  \centerline{\includegraphics[width=\imwidth, height=\imheight]{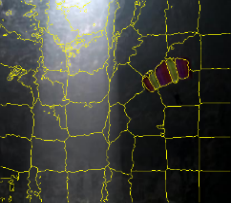}}
\end{minipage}
\caption{Results of 3D SLIC voxel grouping on three consecutive frames. The boundaries of each supervoxel are shown in yellow. The block enclosed by the yellow boundaries in the corresponding position between frames has the same label. (Please print in color.)}
\label{fig:SLIC}
\end{figure}

Note that 3D SLIC does not explicitly enforce connectivity. The adjacency matrix is generated and the clusters with a number of pixels under a threshold are reassigned to the nearest neighbor cluster using connect component analysis. Fig. \ref{fig:SLIC} shows the results of 3D SLIC algorithm after the connect component analysis.

Note that for some HD videos that contain too much redundant details on the background, the SLIC voxel grouping generates some tiny clusters which are too fine and increase the computation and processing time. To solve this problem, it is
recommended to cluster the videos of this kind after the bilateral filtering. The fine edges can be removed and the main boundaries of the object and background would be retained.

\subsection{D. Coarse Segmentation}
\label{ssec:coarse}

\begin{figure}[htb]
\begin{minipage}[b]{.32\linewidth}
  \centering
  \centerline{\includegraphics[width=\imwidth, height=\imheight]{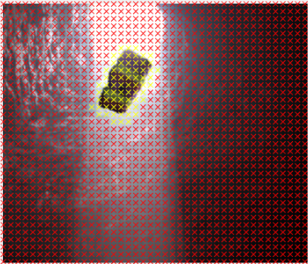}}
  \centerline{(a)}\medskip
\end{minipage}
\begin{minipage}[b]{.32\linewidth}
  \centering
  \centerline{\includegraphics[width=\imwidth, height=\imheight]{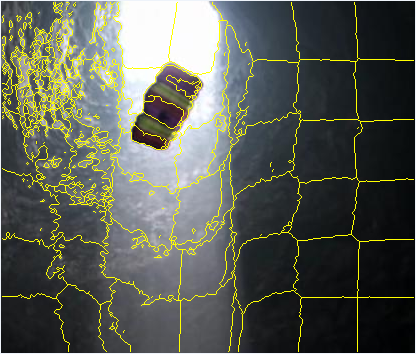}}
  \centerline{(b)}\medskip
\end{minipage}
\begin{minipage}[b]{.32\linewidth}
  \centering
  \centerline{\includegraphics[width=\imwidth, height=\imheight]{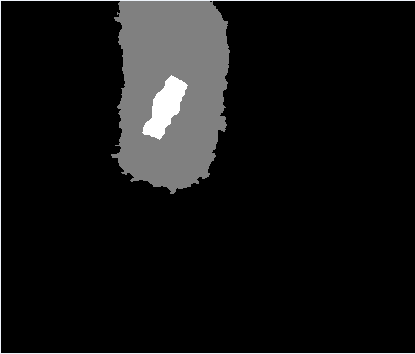}}
  \centerline{(c)}\medskip
\end{minipage}
  \vspace{-.1in}
\caption{Coarse segmentation by combining the results of SSC and 3D SLIC algorithms. (a) Tracking points generated by KLT and SSC. The yellow and red markers represent the foreground and background region respectively; (b) The 3D SLIC supervoxels on the same frame; and (c) The mask generated by combining (a) and (b). The black, gray and white regions denote determined background, undetermined region and determined foreground respectively. (Please print in color.)}
\label{fig:coarse}
\end{figure}

For each supervoxel, coarse segmentation is performed by combining the SSC output and supervoxels. As shown in Fig. \ref{fig:coarse}(a), the SSC algorithm provides an approximate region containing the object of interest. Based on that, we propose a strategy with the following rules: for each supervoxel in the video clip (as shown in Fig. \ref{fig:coarse}(b)), if all the tracking points in it are marked red, this supervoxel is considered as background (black region in Fig. \ref{fig:coarse}(c)); similarly, if all the tracking points in a supervoxel are marked yellow, this supervoxel is labelled as foreground (white region in Fig. \ref{fig:coarse}(c)); otherwise, for the supervoxels containing both colored markers, they are considered as undetermined regions, as shown by the gray region in Fig. \ref{fig:coarse}(c)).

\subsection{E. Graph-based Fine Segmentation}
\label{ssec:fine}

For fine segmentation, we propose to use the GrabCut \cite{Rother_04} algorithm since it requires a set of pixels for background, i.e., it allows incomplete labeling. Also, GrabCut looks for the minimum iteratively rather than in an one-time manner. Each iteration improves the parameters of the GMMs to generate a better segmentation.

\begin{figure}[htb]
\begin{minipage}[b]{.48\linewidth}
  \centering
  \centerline{\includegraphics[width=.9\linewidth]{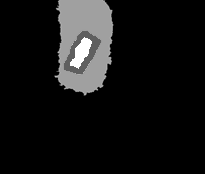}}
  \centerline{(a)}\medskip
\end{minipage}
\begin{minipage}[b]{.48\linewidth}
  \centering
  \centerline{\includegraphics[width=.9\linewidth]{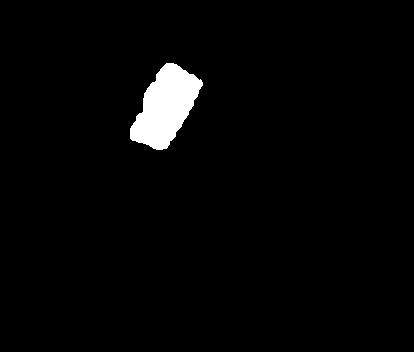}}
  \centerline{(b)}\medskip
\end{minipage}
  \vspace{-.1in}
\caption{Result of fine segmentation using GrabCut method. (a) The algorithm segment the undetermined region to light and dark gray regions; and (b) The light and dark gray regions are merged to the background and foreground respectively to form the final mask.}
\label{fig:fine}
\end{figure}

For the video frames in RGB color space, the object and background are modeled by a full-covariance Gaussian mixture with $K$ components (typically $K = 5$). In order to deal with the GMM tractably, in the optimization framework, an additional vector $\boldsymbol{k}=[k_1,k_2,\cdots,k_n,\cdots,k_N]$ is introduced, with $k_n\in {1,\cdots,K}$, assigning, to each pixel, a unique GMM component, with one component either from the background or the foreground model. Using the mask generated by the coarse segmentation, the black, white and gray regions are flagged with background, foreground and undetermined, or simply marked as 0, 1 or 2 for the image. Applying $k$-means clustering, the pixels belonging to either object or background are clustered into $K$ groups (GMMs). The mean and covariance of the GMM can be estimated by the RGB values of pixels in each cluster, and the weight can be determined by the ratio of the number of pixels in the cluster to the number of overall pixels. Finally, use texture (color) and boundary (contrast) information to get a reliable segmentation result within a few iterations, as illustrated in Fig. \ref{fig:fine}.

\section{Experimental Results}
\label{sec:results}

We conduct experiments on a variety of video content. We run the proposed algorithm on multiple types of data, and generate a mask of the extracted object for each frame. We also compare our segmentation results to those produced by other state-of-the-art methods \cite{D_Zhang_13,Chockalingam_09,Lee_11,Ma_12,Tsai_12}. Both qualitative and quantitative results will be presented to support the effectiveness and robustness of our proposed method.

\subsection{A. Parameters Settings}
The parameters used in the experiments are listed as follows. In the point tracking and clustering process, we set the initial point sampling interval as 10 pixels and the tracking points are reset every 5 frames. The number of clustering groups depends on the application. Typically, we set it to 5. To group pixels, the 3D SLIC algorithm is performed every 30 frames (clip size). For demonstration, the desired number of supervoxels in one frame is set to 100; the desired depth of supervoxels is $D=5$ frames; the regularity $m = 22$; depth of supervoxel $D=5$; the weights for temporal distance and $L^*$ channel are $w_Z=50$ and $w_L=1$ respectively. On average, the 3D SLIC algorithm runs 5 iterations to get a reliable result. To construct the visual effects, the brightness, size, transparency and location of the extracted object and the background image/video could be adjusted and controlled by the user control.

\subsection{B. Evaluation on SegTrack Dataset}
\label{ssec:SegTrack}

We first consider video sequences from the SegTrack \cite{Tsai_12} dataset since a pixel-level segmentation ground truth for each video is available. To quantitatively evaluate the segmentation performance, we use the ground truth provided with the original data. We compare our method with five state-of-the-art methods as shown in Table \ref{tab:error}. The ``penguin'' video sequence is not available for our segmentation application since the ground truth for the ``penguin'' sequence is designed for object tracking in a weakly supervised setting, in which only one penguin is manually annotated by the original user at the each frame. Note that our method is an unsupervised methods, whereas \cite{Chockalingam_09} and \cite{Tsai_12} are supervised method which needs an initial annotation for the first frame. One can see that our algorithm outperforms the other unsupervised methods except for the ``parachute'' and ``birdfall2'' video where it is still comparable to the best one. As mentioned before, for the ``parachute'' video sequence, our result is based on the fact that the person under the parachute should be a part of the object and extracted. However, the person in the ground truth of ``parachute'' sequence was removed in the original dataset, which leads to a slightly inaccurate error calculation. Due to the small size of moving object and the complex background in the scene, the pre-defined density of tracking points may not be high enough to extract the foreground in ``birdfall2'' video sequence, which leads to the pixel error a little higher than the best one. However, this can be improved by making the density of the tracking points self-adjustable. The results in Table \ref{tab:error} take the average of the difference between pixel error and the ground truth, i.e.,
\begin{align}
\text{error}=\dfrac{xor(\text{our result, ground truth})}{\text{number of frames}}
\end{align}
where $xor$ is an exclusive OR operation.

\begin{table}[!htb] \centering
    \caption{Table I: Quantitative pixel-level errors and comparison with the state-of-the-art methods on SegTrack dataset. }
    \label{tab:error}
	\begin{center}
	\begin{threeparttable}
    \begin{tabular}{c|ccccc|c}
		\hline
	        & \cite{D_Zhang_13} & \cite{Chockalingam_09} & \cite{Lee_11} & \cite{Ma_12} & \cite{Tsai_12} & Ours	\\
		\hline
parachute & 220  & 502	& 201  & 221  & 235  & 219	\\
girl      & 1488 & 1755 & 1785 & 1698 & 1304 & 1471	\\
monkeydog & 365  & 683  & 521  & 472  & 563 & 345  \\
birdfall2 & 155  & 454  & 288  & 189  & 252 & 232  \\
cheetah   & 633 & 1217  & 905 & 806 & 1142 & 621	 \\
penguin\tnote{*}	 & NA    & NA   & NA   & NA   & NA   & NA  \\
		\hline
    \end{tabular}
    \begin{tablenotes}
        \item[*] The video sequence ``penguin'' is not applicable to this evaluation.
    \end{tablenotes}
	\end{threeparttable}
	\end{center}
\end{table}

\begin{figure}[htb]
\begin{minipage}[b]{.32\linewidth}
  \centering
  \centerline{\includegraphics[width=\imwidth, height=\imheight]{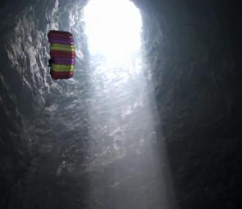}}
\end{minipage}
\begin{minipage}[b]{.32\linewidth}
  \centering
  \centerline{\includegraphics[width=\imwidth, height=\imheight]{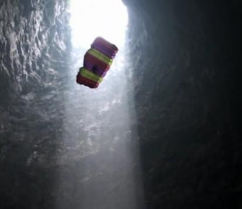}}
\end{minipage}
\begin{minipage}[b]{.32\linewidth}
  \centering
  \centerline{\includegraphics[width=\imwidth, height=\imheight]{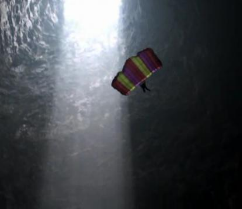}}
\end{minipage}
\centerline{(a) Original frames}\medskip
\vspace{-.05in}  
  
\begin{minipage}[b]{.32\linewidth}
  \centering
  \centerline{\includegraphics[width=\imwidth, height=\imheight]{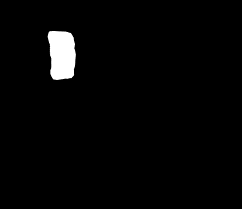}}
\end{minipage}
\begin{minipage}[b]{.32\linewidth}
  \centering
  \centerline{\includegraphics[width=\imwidth, height=\imheight]{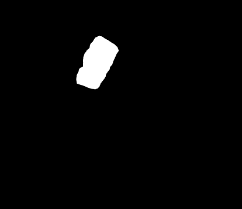}}
\end{minipage}
\begin{minipage}[b]{.32\linewidth}
  \centering
  \centerline{\includegraphics[width=\imwidth, height=\imheight]{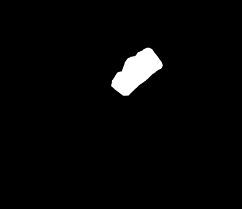}}
\end{minipage}
\centerline{(b) Ground truth}\medskip
\vspace{-.05in}  
  
\begin{minipage}[b]{.32\linewidth}
  \centering
  \centerline{\includegraphics[width=\imwidth, height=\imheight]{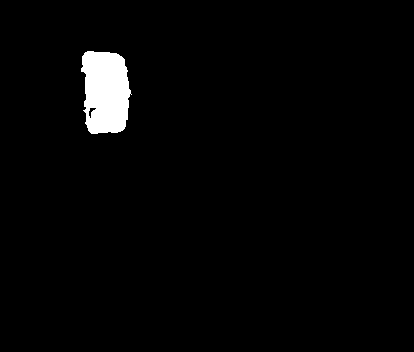}}
\end{minipage}
\begin{minipage}[b]{.32\linewidth}
  \centering
  \centerline{\includegraphics[width=\imwidth, height=\imheight]{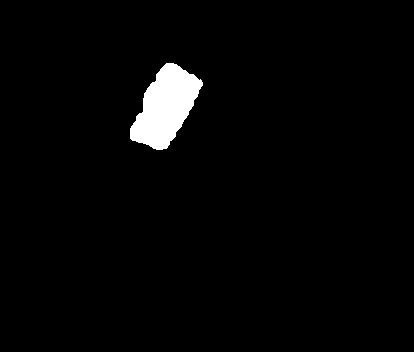}}
\end{minipage}
\begin{minipage}[b]{.32\linewidth}
  \centering
  \centerline{\includegraphics[width=\imwidth, height=\imheight]{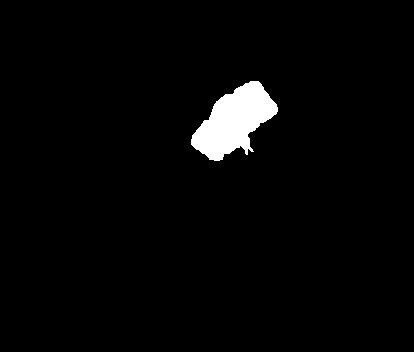}}
\end{minipage}
\centerline{(c) Our results}\medskip
\vspace{-.1in}  
\caption{Qualitative results of SegTrack ``parachute'' video sequence.}
\label{fig:parachute}
\end{figure}

\begin{figure}[htb]
\begin{minipage}[b]{.32\linewidth}
  \centering
  \centerline{\includegraphics[width=1.05in, height=0.84in]{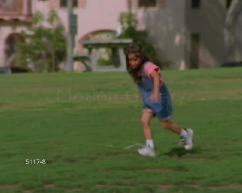}}
\end{minipage}
\begin{minipage}[b]{.32\linewidth}
  \centering
  \centerline{\includegraphics[width=1.05in, height=0.84in]{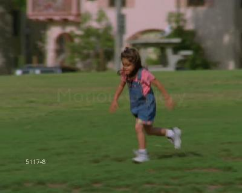}}
\end{minipage}
\begin{minipage}[b]{.32\linewidth}
  \centering
  \centerline{\includegraphics[width=1.05in, height=0.84in]{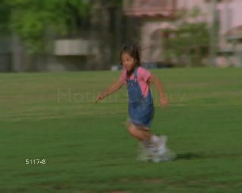}}
\end{minipage}
\centerline{(a) Original frames}\medskip
\vspace{-.05in}  
  
\begin{minipage}[b]{.32\linewidth}
  \centering
  \centerline{\includegraphics[width=1.05in, height=0.84in]{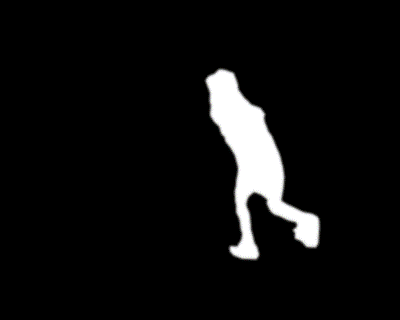}}
\end{minipage}
\begin{minipage}[b]{.32\linewidth}
  \centering
  \centerline{\includegraphics[width=1.05in, height=0.84in]{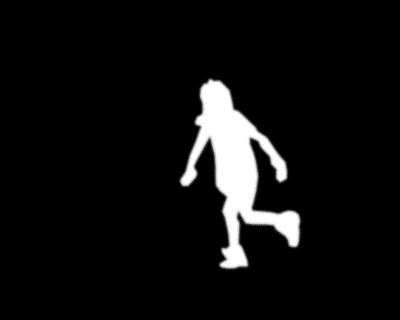}}
\end{minipage}
\begin{minipage}[b]{.32\linewidth}
  \centering
  \centerline{\includegraphics[width=1.05in, height=0.84in]{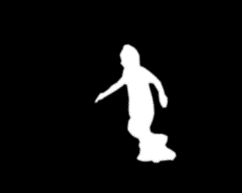}}
\end{minipage}
\centerline{(b) Ground truth}\medskip
\vspace{-.05in}  
  
\begin{minipage}[b]{.32\linewidth}
  \centering
  \centerline{\includegraphics[width=1.05in, height=0.84in]{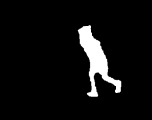}}
\end{minipage}
\begin{minipage}[b]{.32\linewidth}
  \centering
  \centerline{\includegraphics[width=1.05in, height=0.84in]{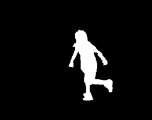}}
\end{minipage}
\begin{minipage}[b]{.32\linewidth}
  \centering
  \centerline{\includegraphics[width=1.05in, height=0.84in]{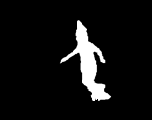}}
\end{minipage}
\centerline{(c) Our results}\medskip
\vspace{-.1in}  
\caption{Qualitative results of SegTrack ``girl'' video sequence.}
\label{fig:girl}
\end{figure}

Fig. \ref{fig:parachute} shows an example of the qualitative results of ``parachute'' video sequence in SegTrack dataset. In this video sequence, the foreground and background regions move in different ways. Compared to the last column of Fig. \ref{fig:parachute}(b) and (c), the person under the parachute is segmented into the foreground in our results instead of merged into background as shown in ground truth. This makes the segmentation result more reasonable, although it leads to the slight error increase in Table \ref{tab:error}. Fig. \ref{fig:girl} compares our results with the ground truth on ``girl'' video sequence. The ``girl'' video sequence suffers from low resolution and severe motion blur which increases the difficulty for segmentation. The point tracking and supervoxel generation are affected by the motion blur. This becomes the main source of the pixel-level error. 

All the experiments are performed on an Intel\textsuperscript{\textregistered} Core\texttrademark i5-4590 CPU at 3.30GHz with 16GB memory. Before extensive code and data structure optimization, the processing time per frame is around 0.52s, 15.86s, and 7.62s for points clustering, supervoxel generation and final segmentation respectively.

\subsection{C. Evaluation on Kodak Alaris Consumer Video Dataset}
\label{ssec:Kodak}

\begin{figure}[htb!]
\begin{minipage}[b]{.48\linewidth}
  \centering
  \centerline{\includegraphics[width=1.42in, height=0.8in]{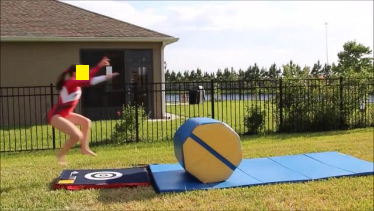}}
\end{minipage}
\begin{minipage}[b]{.48\linewidth}
  \centering
  \centerline{\includegraphics[width=1.42in, height=0.8in]{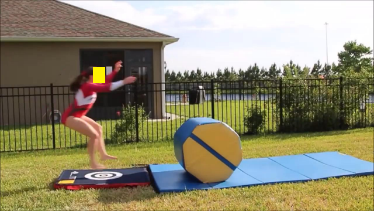}}
\end{minipage}

\centerline{(a) Original frames in the video sequence.}\medskip
\vspace{-.05in}  
  
\begin{minipage}[b]{.48\linewidth}
  \centering
  \centerline{\includegraphics[width=1.42in, height=0.8in]{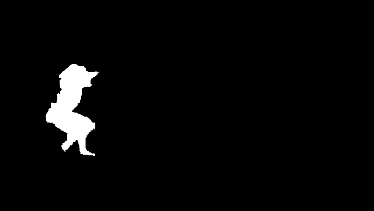}}
\end{minipage}
\begin{minipage}[b]{.48\linewidth}
  \centering
  \centerline{\includegraphics[width=1.42in, height=0.8in]{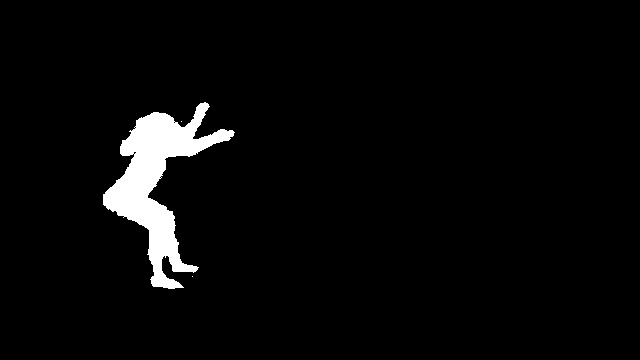}}
\end{minipage}

\centerline{(b) Mask representing the extracted object in the sequence.}\medskip
\vspace{-.1in}  
\caption{Object segmentation results on ``gymnast1'' video sequence in Kodak Alaris consumer video dataset.}
\label{fig:gymnast}
\end{figure}

\begin{figure}[htb!]
\begin{minipage}[b]{.48\linewidth}
  \centering
  \centerline{\includegraphics[width=1.42in, height=0.8in]{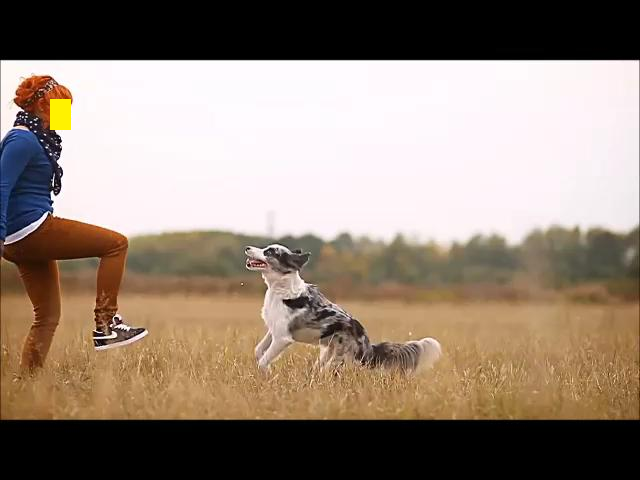}}
\end{minipage}
\begin{minipage}[b]{.48\linewidth}
  \centering
  \centerline{\includegraphics[width=1.42in, height=0.8in]{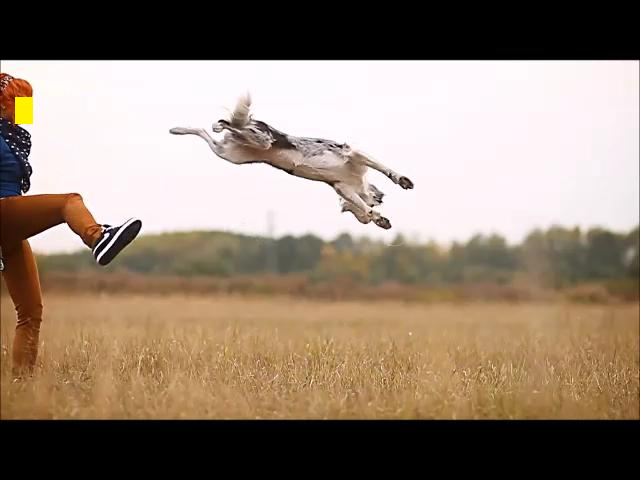}}
\end{minipage}

\centerline{(a) Original frames in the video sequence.}\medskip
\vspace{-.05in}  
  
\begin{minipage}[b]{.48\linewidth}
  \centering
  \centerline{\includegraphics[width=1.42in, height=0.8in]{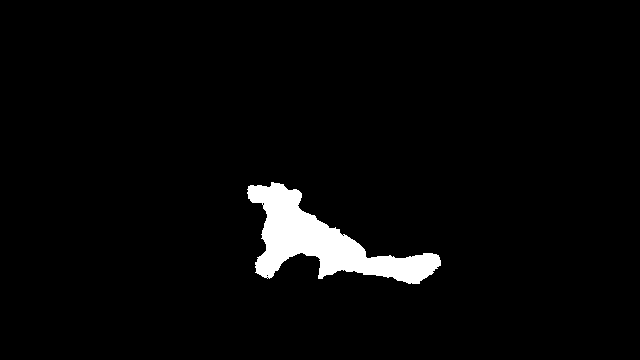}}
\end{minipage}
\begin{minipage}[b]{.48\linewidth}
  \centering
  \centerline{\includegraphics[width=1.42in, height=0.8in]{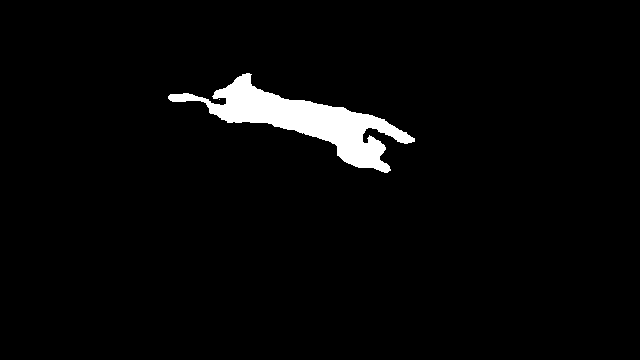}}
\end{minipage}

\centerline{(b) Mask representing the extracted object in the sequence.}\medskip
\vspace{-.1in}  
\caption{Object segmentation results on ``dog'' video sequence in Kodak Alaris consumer video dataset.}
\label{fig:dog}
\end{figure}

With the rapid development and lower cost of smartphones and new digital capture devices, consumer videos are becoming ever popular as is evident by the large volume of YouTube video upload, as well as video viewing in Facebook social network. These large amount of videos also pose a challenge for organizing and retrieving videos for the consumers. Besides the SegTrack dataset, we have conducted evaluations of our proposed approach on some of the videos from Kodak Alaris consumer video dataset. The videos in the dataset are mostly captured in standard HD format with high frame resolution. Fig. \ref{fig:gymnast} shows the qualitative results of ``gymnast1'' video sequence in this dataset. Because of the high resolution of the video, we apply bilateral filtering on the original frame to remove some fine details of the background and keep the main edges. The bilateral filtering does not affect the performance of either SSC or 3D SLIC algorithm, but rather saves the computation. Another example is shown in Fig. \ref{fig:dog}. In this video, some parts of the moving object (dog) is similar to the background trees in color, and the other parts are as white as the background sky. It turns out that our algorithm produces reasonably good results for this difficult task.

\section{Conclusion}
\label{sec:conclusions}

We have proposed a novel and accurate coarse-to-fine approach to segment the salient object in video sequences. This approach involves a parallel scheme, which consists of KLT, SSC and 3D SLIC algorithm to identify the approximate location of the most salient object. Subsequently, an unsupervised graph-based method is used for fine segmentation. Since the coarse segmentation determines the location of the moving object rather than the exact boundaries, the robustness of this approach can be guaranteed. It is also worth mentioning that this algorithm can be easily extended to multiple objects segmentation by controlling the number of classes in the SSC stage. Compared to other state-of-the-art approaches, it has stronger ability to segment video sequences accurately in any resolution and length within a shorter time. The experimental results also validate the effectiveness and performance of the proposed method.


\small
\bibliographystyle{IEEEbib}
\bibliography{refs}


\begin{biography}
\textbf{Chi Zhang} received his MS in electrical engineering from Rochester Institute of Technology (2013) and is currently a Ph.D. student in imaging science at Rochester Institute of Technology. He had worked as an software intern at Kodak Alaris Inc. in Rochester, NY. In recent years his professional interests focus on the area of computer vision, including image analysis, video processing and convolutional neural networks for visual recognition. He is a student member of IEEE.

\textbf{Alexander Loui} received his Ph.D. (1990) in Electrical Engineering from the University of Toronto, Canada. He is currently a Senior Principal Scientist at Kodak Alaris in Rochester, NY. He is also an Adjunct Professor of ECE Department at Ryerson University. Dr. Loui has been directing research on multimedia processing, video analysis and summarization, image management and retrieval, event detection, image quality assessment, and computer vision applications. He is a Fellow of IEEE and SPIE.

\end{biography}

\end{document}